\documentclass[10pt, conference, compsocconf]{IEEEtran}
\IEEEoverridecommandlockouts
\IEEEpubid{\makebox[\columnwidth]{\copyright2018 IEEE \hfill} \hspace{\columnsep}\makebox[\columnwidth]{ }}

\usepackage[utf8]{inputenc}
\usepackage{url}
\usepackage[bookmarks, bookmarksopen, bookmarksnumbered, colorlinks,linkcolor=blue,urlcolor=blue,citecolor=blue]{hyperref}
\usepackage{cite}
\usepackage{amsmath,amssymb,amsfonts}
\usepackage{graphicx}
\usepackage{xcolor}

\usepackage{adjustbox}
\usepackage{cuted}
\usepackage{capt-of}
\usepackage{lscape}
\usepackage{balance}

\ifCLASSINFOpdf
\else
\fi
\begin{document}
%
\title{Mapping the research software sustainability space}


\author{\IEEEauthorblockN{Stephan Druskat}
\IEEEauthorblockA{Dept. of German Studies and Linguistics\\
Humboldt-Universität zu Berlin\\
Berlin, Germany\\
stephan.druskat@hu-berlin.de}
\and
\IEEEauthorblockN{Daniel S. Katz}
\IEEEauthorblockA{NCSA, CS, ECE, iSchool\\
University of Illinois Urbana-Champaign\\
Urbana, IL, USA\\
d.katz@ieee.org}
}


%


\maketitle
\IEEEpubidadjcol

\begin{abstract}
A growing number of largely uncoordinated initiatives focus on research software sustainability. 
A comprehensive mapping of the research software sustainability space can help identify gaps in their
efforts, track results, and avoid duplication of work.
To this end, this paper suggests enhancing an existing schematic of activities in research software
sustainability, and formalizing it in a directed graph model.
Such a model can be further used to define a classification schema which, applied to research results in the
field, can drive the identification of past activities and the planning of future efforts.

\end{abstract}


%
\IEEEpeerreviewmaketitle

The number of activities with a focus on the sustainability of research
software has increased over the last few years. Its major proponents are the
Software Sustainability Institute (\href{https://software.ac.uk/}{software.ac.uk})
(SSI), founded in 2010 in the UK, the international WSSSPE
organization (\href{http://wssspe.researchcomputing.org.uk/}{wssspe.researchcomputing.org.uk}) with its
workshop series and working groups, and to a certain extent, the
international community of Research Software Engineers (\href{https://rse.ac.uk/community/international-rse-groups/}{rse.ac.uk/community/international-rse-groups}),
which also originated in the UK. This trend is ongoing, as the planning for an institution similar to the SSI in the US, the US
Research Software Sustainability Institute (URSSI, \href{http://urssi.us}{urssi.us}), and the recent formation of the Better Scientific Software
community (BSSw, \href{https://bssw.io/}{bssw.io}) show. Additionally, further
entities are active in the research software sustainability space (``the
space''), on different levels. These include, for example, working groups,
special interest groups, and others, on local, institutional, or
disciplinary levels.

All the above-mentioned entities conduct research on the sustainability
of research software, and publish in different venues, e.g., through
papers, blog posts, as well as talks and presentations at workshops and
conferences. Their research feeds back into academic institutions and
associations on the policy and education levels, and increasingly with
regard to human resources. Their findings also inform educational
activities such as The Carpentries (\href{https://carpentries.org/}{carpentries.org}), 
which teaches foundational coding to researchers, including
specialized communities, such as the high performance computing and
digital humanities communities.

These efforts are not currently formally coordinated, and
it is unclear whether formal -- or even central -- coordination
is at all useful, desirable, or achievable. After all, the
dynamic nature of, e.g., small working groups, workshops, and spontaneous
events such as hack days are often more constructive and productive than
larger, coordinated efforts. The autonomy of activities
bears the risk of unnecessary duplication of efforts, as well as neglect
of specific areas of the space.

\section{Mapping the space}

A comprehensive mapping of the research software sustainability space
can help to reduce this risk and support further efforts.

Existing classification schemes, such as the ACM's Computing Classification System
(\href{https://www.acm.org/about-acm/class}{acm.org/about-acm/class}), are
unsuitable for such a mapping, as they do not reflect the specific configuration
of this domain in terms of agents and activities.

\begin{figure}[htbp]
\centerline{\includegraphics[width=\columnwidth]{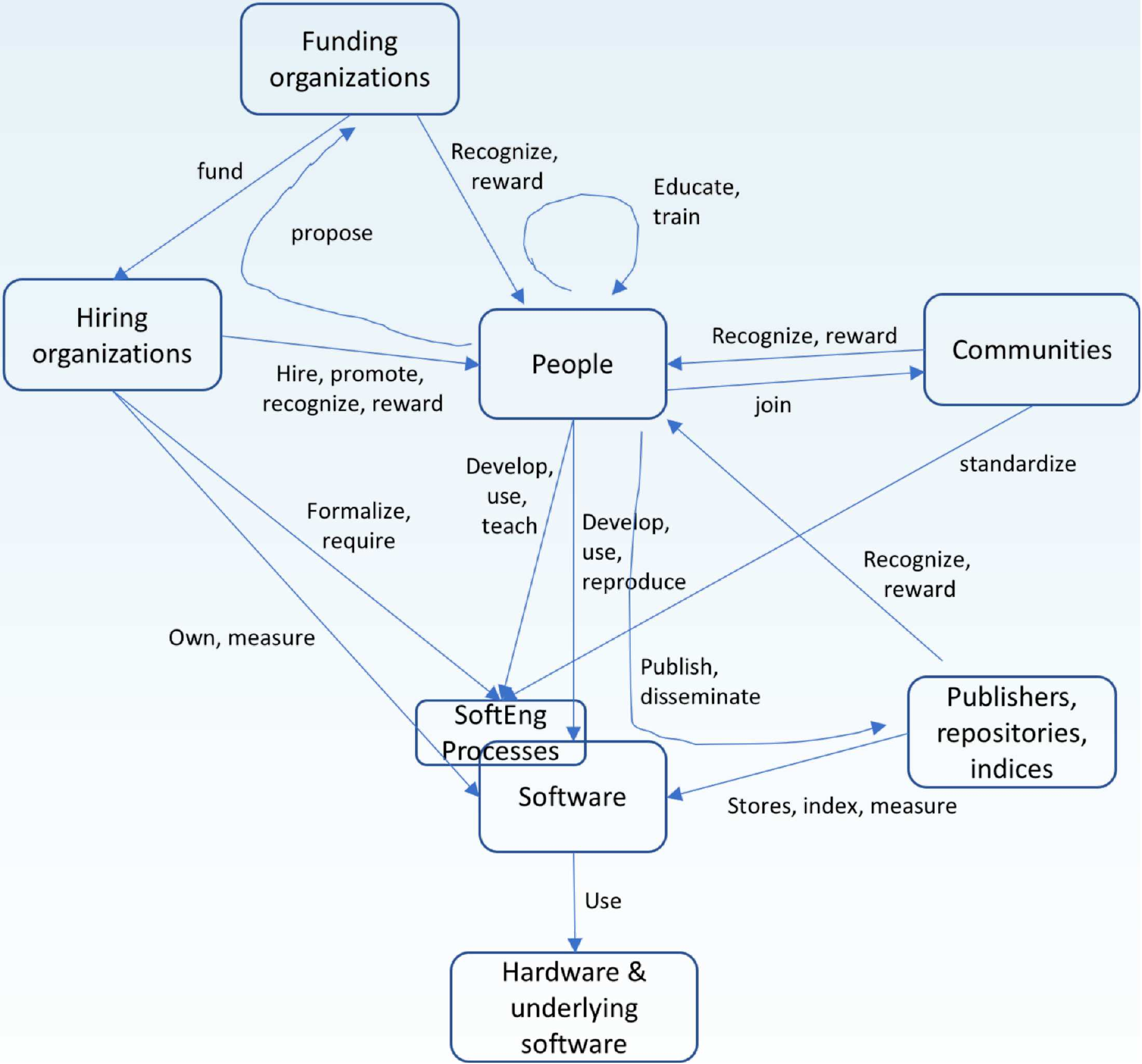}}
\caption{Reproduction of Katz' original sustainability schematic \cite[p. 3]{katz_research_2018}.}
\label{fig:katz-orig}
\end{figure}

A first mapping was introduced by Katz \cite{katz_research_2018} in a
directed graph schematic, reproduced in Figure \ref{fig:katz-orig}. This
visualization lists important parties (nodes) and activities (edges) in
the space.

\begin{figure*}[htbp]
\centerline{\includegraphics[width=\textwidth]{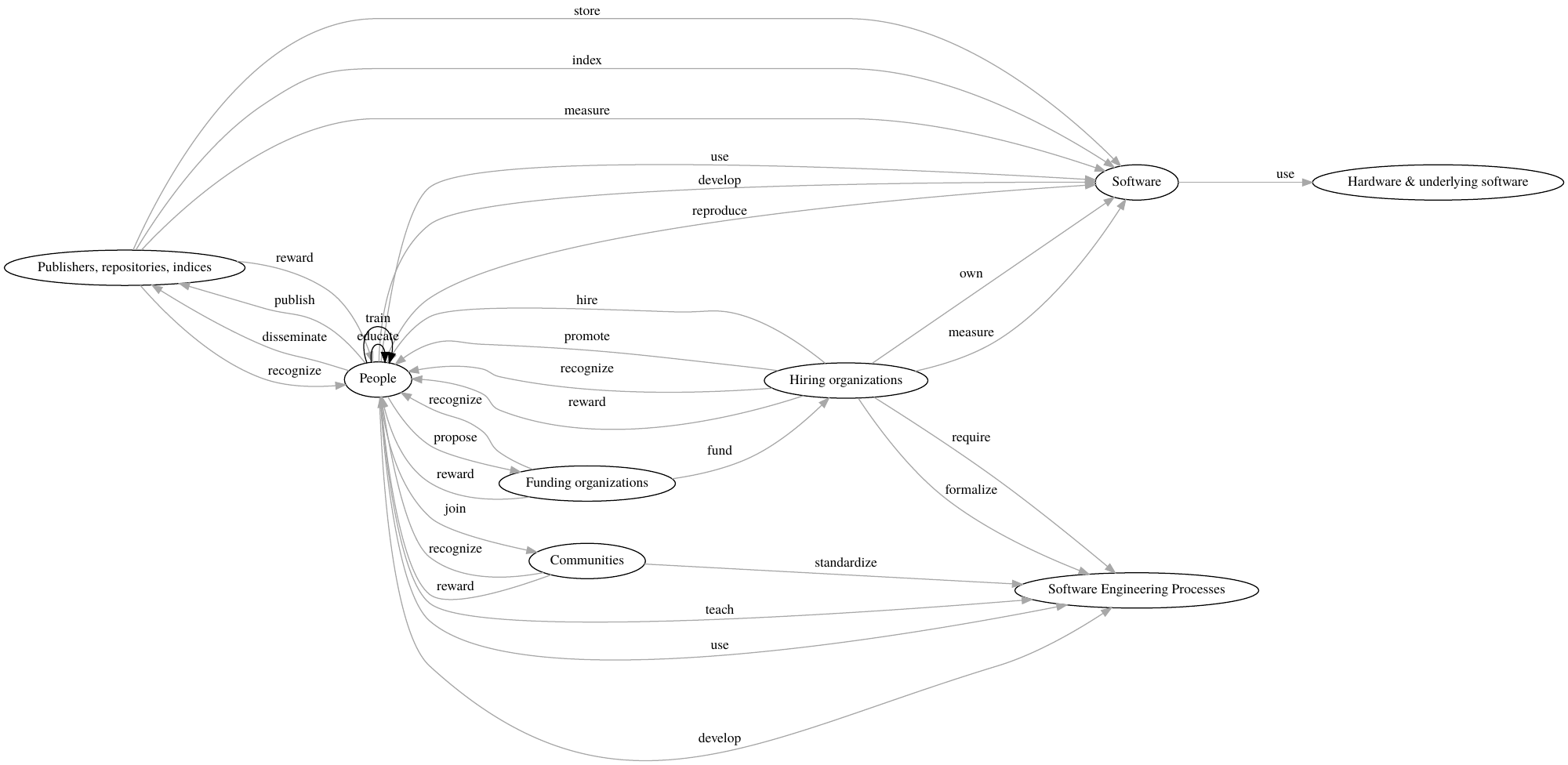}}
\caption{Visualization of the formalized resosuma model of Katz' sustainability schematic.}
\label{fig:graph-katz}
\end{figure*}

Based on this initial concept, a more comprehensive mapping can be built.
Advances may include disassembling combined activities, a
higher resolution and more precise definition of involved parties, as
well as adding potentially missing agents.

Additionally, formalizing the mapping can make it useful for further
processing and would benefit the community: it enables different efforts in the space to be classified based on the
parties and activities they cover. This includes different possible
formats, such as research outputs (papers, blog posts, talks), events,
groups, and projects. Such a classification in turn enables the
gaps in community efforts to be identified via quantitative methods,
e.g., publication and event analysis. These gaps can then be closed by
initiating new efforts, e.g., workshops and projects in these areas.

\subsection{Initial formalization}\label{initial-formalization}

In a first step towards a formalization, we mapped Katz' original sustainability
schematic (Figure \ref{fig:katz-orig}) to a \emph{resosuma}
CSV representation of activities \cite{stephan_druskat_research-software/resosuma-data_2018}. In the
process, activities that are combined in the schematic (i.e., edge
labels that are lists of verbs) were disassembled, so that each verb
(the \emph{action}) has its own row, with its subject (the \emph{actor})
in the cell to its left and its object (the \emph{actee}) in the cell
to its right.

Figure~\ref{fig:graph-katz} shows a visualization of the data, created with the
\emph{resosuma}\cite{druskat_resosuma_2018-1} Python package.

\subsection{Refinement}\label{refinement}

In a second step, we made some changes to the data to more
comprehensively represent activities in the research software
sustainability space, at a higher resolution \cite{resosumadata4}.

The changes have been introduced based on introspection alone, and in order to
reach a sufficient level of comprehensiveness, future refinement work towards
a stable version of the model should include a structured literature
analysis.

The respective \emph{resosuma} visualization is given in Figure
\ref{fig:graph-resosuma-0.2.0}. The changes included the following.

\textbf{Differentiating the ``People'' node into ``Research Software
Engineers'', ``Researchers'' and ``Educators'' nodes, and defining
related activities respectively.} This differentiates
between groups of \emph{roles} that have distinct scopes of
responsibilities within the space. While ``Research Software Engineers''
are active in producing and publishing research software and applying
``Software Engineering Processes'', ``Researchers'' are active in using
research software and publishing research products (but not research
software), but have little direct stake in ``Software Engineering
Processes''. These two roles rely on different requirements to be
fulfilled within the space. Similarly, ``Educators'', including
academic staff, Carpentry instructors, and training consultants, do not
use research software in this role, but disseminate ``Software
Engineering Processes'' for example.

\textbf{Differentiating the ``Publishers, repositories, indices''
node into separate nodes, and defining related activities
respectively.} We split these parties over separate nodes because they play differing roles in the space. Additionally, we split
``repositories'' into ``code platforms'' and
``archival repositories''. ``Code platforms'' serve the purpose of
interacting with code, allow operations like forking, branching,
and often provide further functionality for collaboration. Well-known
examples include GitHub (\href{https://github.com}{github.com}), GitLab (\href{https://gitlab.com}{gitlab.com})
and Bitbucket (\href{https://bitbucket.org/}{bitbucket.org}), but ``Code platforms'' may also include simpler version control system repositories. ``Archival
repositories'' archive and publish versions of research products including software,
but are usually not interactive. Examples include general purpose data
repositories such as Zenodo (\href{https://zenodo.org/}{zenodo.org}) or
figshare (\href{https://figshare.com/}{figshare.com}), build artifact repositories
such as The Central Repository (``Maven Central'', \href{https://search.maven.org/}{search.maven.org}),
the Python Package Index (PyPI, \href{https://pypi.org}{pypi.org}), and others,
e.g., preprint repositories such as arXiv.org (\href{https://arxiv.org/}{arxiv.org}).

\textbf{Differentiating the ``Software'' node into ``Research
software'' and ``Supporting software'' nodes, and defining related
activities respectively.} While the differentiated nodes represent similar objects, the ways they are produced and used, 
and the ways they are referenced and rewarded differ remarkably. Additionally, research
software is the defining entity at the core of the space, and
``supporting software'' can exist independently of it while potentially having
a great effect on the space. ``Supporting software'' covers
software that is used to create research software, including everything from
editors and integrated development environments to version control
system platforms.

\textbf{The ``Software Engineering Processes'' node} was left unchanged,
but it should be noted that it includes a wide range of processes and
practices, including best practices around metadata, citation, etc.

\textbf{``are''} has been introduced as a specific relation in the
space, where ``Research Software Engineers'' and ``Researchers'' are
nodes that may reference different roles for the same individual. As
this relation is at the heart of some of the activities in the space,
most notably the international RSE community, it should be explicitly
included in the mapping.

\textbf{Further activities were added.} We established an ``own'' activity, for
example, from ``Research Software Engineers'' to
``Research Software'' in addition to the one from ``Hiring
organizations'' to ``Research Software'' to represent potential
copyright of an RSE on their work (cf. \cite{forgo_legal_2017}).

A visualization of the refined resosuma data is given
in Table \ref{table:resosuma-v2} and Figure \ref{fig:graph-resosuma-0.2.0}.
A more optimal visualization of this data would be an interactive, zoomable graph
visualization in a web application.

\section{Future work}\label{future-work}

The data is openly maintained on GitHub
(\href{https://github.com/research-software/resosuma-data}{github.com/research-software/resosuma-data}),
and we propose that the community should collaborate to

\begin{enumerate}
\def\labelenumi{\arabic{enumi}.}
\item
 complete it to create iterative releases which reflect state changes
 in the research software sustainability space; and
\item
 build a classification scheme on it that can be used to classify
 efforts.
\end{enumerate}

The classification could take the form of simple handles for activities,
e.g., \emph{resosuma:rse-dev-rso} for the activity ``Research Software
Engineers develop research software''. The classification could then be
used to tag efforts, retrospectively where possible. This is easily done
for publications, where classification handles can be included in
keywords or in the body, but should also be done for projects, working
groups, events, etc., for example on their websites.

The classification could be used for automated extraction and quantitative
analysis, and could also serve to build a registry of efforts across
formats that would make it even easier to identify gaps and track
progress.

Additionally, the classification could be used by funders and
institutions to analyze their portfolio for reporting and planning, similar
to the NIH's (\href{https://www.nih.gov/}{nih.gov})
use of the ``Research, Condition, and Disease Categorization Process'' (RCDC, \href{https://report.nih.gov/rcdc/}{report.nih.gov/rcdc/}).

\begin{landscape}
\begin{figure}
   \centering
    \includegraphics[height=0.95\textheight]{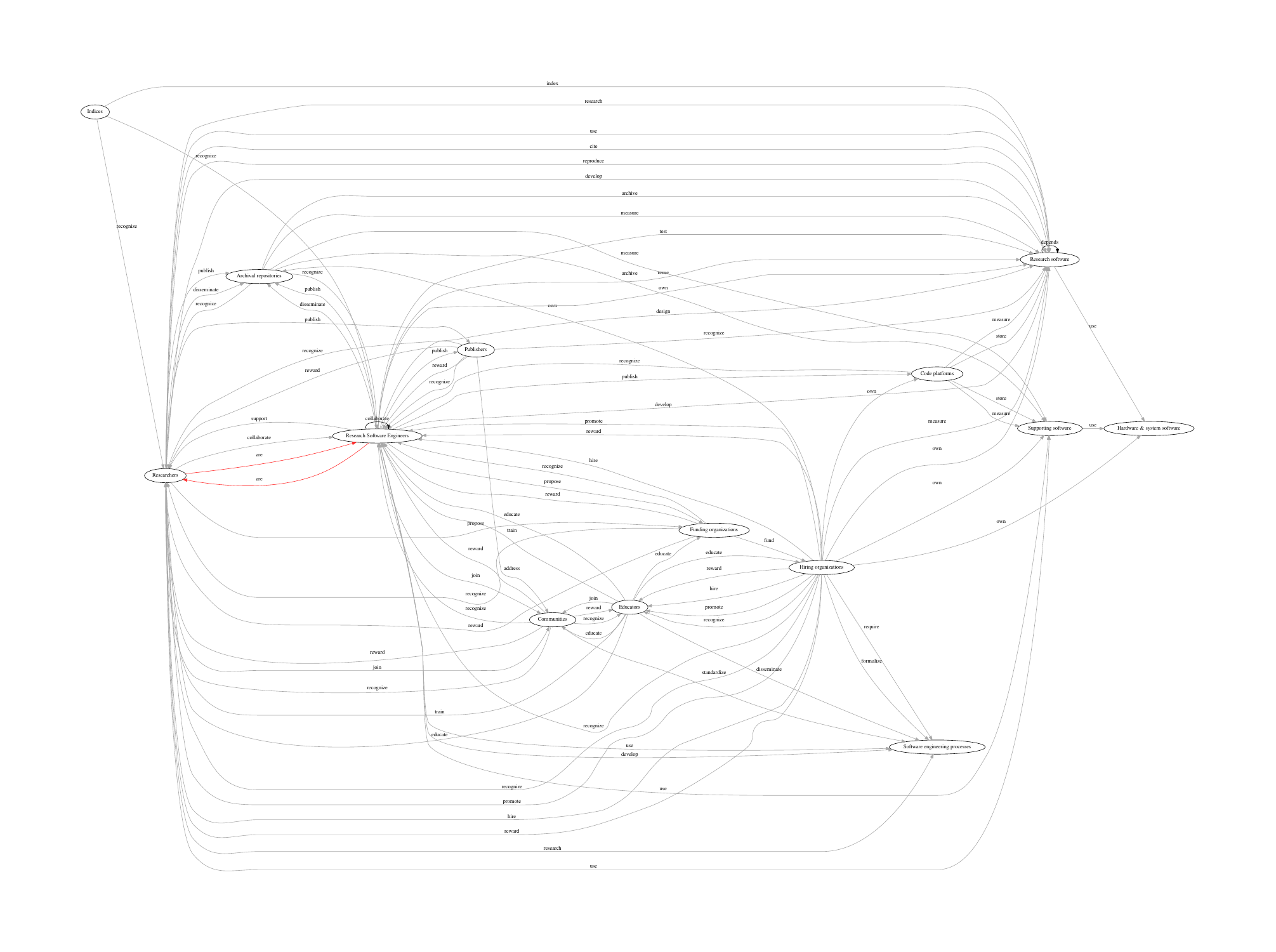}
    \caption{Visualization of the refined resosuma model \cite{resosumadata4}.}
    \label{fig:graph-resosuma-0.2.0}
\end{figure}
\end{landscape}

The level of granularity of the activity graph should be discussed
within the community, and if deemed necessary and helpful, different
resolutions of the mapping could be created, e.g., a low-res version
that would look similar to Figure \ref{fig:graph-katz}, and a high-res
version that would look similar to an optimized version of Figure
\ref{fig:graph-resosuma-0.2.0}. Both versions can be codified in different
versions of the classification scheme, similarly to what has been done in
ISO 639 (\href{https://en.wikipedia.org/wiki/ISO\_639}{en.wikipedia.org/wiki/ISO\_639}), with a low-res
classification in ISO 639-1 and a more high-res classification in ISO
639-3.

Future refinement work towards a stable version of the model should also
include a structured review of the available literature and past activities.

To promote use of the classification, conferences, workshops and similar 
events, and editors of publications in the research software sustainability space should ask for
contributions to include the applicable resosuma handles in their keywords sections, once these handles have 
been developed. Alternatively, reviewers of these contributions could be asked to classify them during the
review process.

Additionally, researchers should be encouraged to add resosuma handles that cover their expertise
to their public profiles, e.g., their ORCID (\href{https://orcid.org}{orcid.org}) profiles. This
will enable easier identification of potentially suitable reviewers for future contributions
within the space.

To facilitate access to the handles and the map itself, and to accumulate future literature
and activity analyses, the community should develop a central resource for resosuma, e.g., a community-curated
website.
Such a resource should also include a more accessible visualization than has been possible here in Figure
\ref{fig:graph-resosuma-0.2.0} and Table \ref{table:resosuma-v2}, e.g., in an interactive webpage.

\begin{strip}
\captionof{table}{Activities modeled in \cite{resosumadata4}.}\label{table:resosuma-v2}
\begin{center}
\begin{adjustbox}{max width=\textwidth}
\begin{tabular}{|rcl|rcl|}
\hline
\textbf{Actor}  & \textbf{Action} & \textbf{Actee} & \textbf{Actor} & \textbf{Action} & \textbf{Actee} \\ \hline
Research Software Engineers & collaborate & Research Software Engineers& Code platforms & measure & Research software \\ \hline
Research Software Engineers & support & Researchers& Code platforms & store  & Supporting software\\ \hline
Research Software Engineers & are & Researchers& Code platforms & measure & Supporting software\\ \hline
Research Software Engineers & publish & Publishers & Archival repositories & recognize  & Research Software Engineers\\ \hline
Research Software Engineers & publish & Code platforms & Archival repositories & recognize  & Researchers\\ \hline
Research Software Engineers & publish & Archival repositories & Archival repositories & archive & Research software \\ \hline
Research Software Engineers & disseminate & Archival repositories & Archival repositories & measure & Research software \\ \hline
Research Software Engineers & develop & Research software & Archival repositories & archive & Supporting software\\ \hline
Research Software Engineers & test& Research software & Archival repositories & measure & Supporting software\\ \hline
Research Software Engineers & design & Research software & Indices  & index  & Research software \\ \hline
Research Software Engineers & reuse  & Research software & Indices  & recognize  & Research Software Engineers\\ \hline
Research Software Engineers & own & Research software & Indices  & recognize  & Researchers\\ \hline
Research Software Engineers & use & Supporting software& Research software & depends & Research software \\ \hline
Research Software Engineers & join& Communities& Research software & use& Hardware \& system software\\ \hline
Research Software Engineers & propose & Funding organizations & Supporting software  & use& Hardware \& system software\\ \hline
Research Software Engineers & use & Software engineering processes & Communities  & recognize  & Research Software Engineers\\ \hline
Research Software Engineers & develop & Software engineering processes & Communities  & reward & Research Software Engineers\\ \hline
Researchers & collaborate & Research Software Engineers& Communities  & recognize  & Researchers\\ \hline
Researchers & are & Research Software Engineers& Communities  & reward & Researchers\\ \hline
Researchers & publish & Publishers & Communities  & recognize  & Educators \\ \hline
Researchers & publish & Archival repositories & Communities  & reward & Educators \\ \hline
Researchers & disseminate & Archival repositories & Communities  & standardize & Software engineering processes \\ \hline
Researchers & use & Research software & Funding organizations & recognize  & Research Software Engineers\\ \hline
Researchers & develop & Research software & Funding organizations & reward & Research Software Engineers\\ \hline
Researchers & reproduce  & Research software & Funding organizations & recognize  & Researchers\\ \hline
Researchers & cite& Research software & Funding organizations & reward & Researchers\\ \hline
Researchers & research& Research software & Funding organizations & fund& Hiring organizations  \\ \hline
Researchers & use & Supporting software& Hiring organizations & hire& Research Software Engineers\\ \hline
Researchers & join& Communities& Hiring organizations & promote & Research Software Engineers\\ \hline
Researchers & propose & Funding organizations & Hiring organizations & recognize  & Research Software Engineers\\ \hline
Researchers & research& Software engineering processes & Hiring organizations & reward & Research Software Engineers\\ \hline
Educators  & train  & Research Software Engineers& Hiring organizations & hire& Researchers\\ \hline
Educators  & educate & Research Software Engineers& Hiring organizations & promote & Researchers\\ \hline
Educators  & train  & Researchers& Hiring organizations & recognize  & Researchers\\ \hline
Educators  & educate & Researchers& Hiring organizations & reward & Researchers\\ \hline
Educators  & educate & Communities& Hiring organizations & hire& Educators \\ \hline
Educators  & join& Communities& Hiring organizations & promote & Educators \\ \hline
Educators  & educate & Funding organizations & Hiring organizations & recognize  & Educators \\ \hline
Educators  & educate & Hiring organizations  & Hiring organizations & reward & Educators \\ \hline
Educators  & disseminate & Software engineering processes & Hiring organizations & own & Code platforms \\ \hline
Publishers & reward & Research Software Engineers& Hiring organizations & own & Archival repositories \\ \hline
Publishers & recognize  & Research Software Engineers& Hiring organizations & own & Research software \\ \hline
Publishers & reward & Researchers& Hiring organizations & measure & Research software \\ \hline
Publishers & recognize  & Researchers& Hiring organizations & own & Supporting software\\ \hline
Publishers & recognize  & Research software & Hiring organizations & formalize  & Software engineering processes \\ \hline
Publishers & address & Communities& Hiring organizations & require & Software engineering processes \\ \hline
Code platforms  & recognize  & Research Software Engineers& Hiring organizations & own & Hardware \& system software \\ \hline
Code platforms  & store  & Research software &&& \\ \hline
\end{tabular}
\end{adjustbox}
\end{center}
\end{strip}

\section*{Acknowledgments}

The authors would like to thank Neil Chue Hong for early feedback and contributions to the data model.

S. Druskat would like to acknowledge funding assistance from the Software Sustainability Institute. The Software Sustainability Institute is supported by the EPSRC, BBSRC and ESRC Grant EP/N006410/1.

\bibliographystyle{IEEEtran}
\bibliography{wssspe6-1-paper}

\end{document}